\documentclass[aps,prb,twocolumn,floatfix,superscriptaddress,showpacs,showkeys,amssymb]{revtex4}
\usepackage{graphicx}

\begin{document}

\title{Non-perturbative electron dynamics in crossed fields}
\author{J.~M.~Villas-B\^{o}as}
\affiliation{Department of Physics and Astronomy, Condensed Matter and Surface Science
Program, \\
Ohio University, Athens, Ohio 45701-2979}
\affiliation{Departamento de F\'{\i}sica, Universidade Federal de S\~{a}o Carlos,
13565-905, S\~{a}o Carlos, S\~{a}o Paulo, Brazil}
\author{Wei Zhang}
\author{Sergio E.~Ulloa}
\affiliation{Department of Physics and Astronomy, Condensed Matter and Surface Science
Program, \\
Ohio University, Athens, Ohio 45701-2979}
\author{P.~H.~Rivera}
\altaffiliation[Current address: ]{Consejo Superior de Investigaciones, Universidad Nacional Mayor de San
Marcos, Lima, Peru.}
\author{Nelson Studart}
\affiliation{Departamento de F\'{\i}sica, Universidade Federal de S\~{a}o Carlos,
13565-905, S\~{a}o Carlos, S\~{a}o Paulo, Brazil}
\date{\today }

\begin{abstract}
Intense AC electric fields on semiconductor structures have been studied in
photon-assisted tunneling experiments with magnetic field applied either
parallel ($B_{\parallel }$) or perpendicular ($B_{\perp }$) to the interfaces.
We examine here the electron dynamics in a double quantum well when intense AC
electric fields $F$, and \emph{tilted} magnetic fields are applied
\emph{simultaneously}. The problem is treated \emph{non-perturbatively} by a
time-dependent Hamiltonian in the effective mass approximation, and using a
Floquet-Fourier formalism. For $B_{\parallel}=0$, the quasi-energy spectra show
two types of crossings: those related to different Landau levels, and those
associated to \emph{dynamic localization} (DL), where the electron is confined
to one of the wells, despite the non-negligible tunneling between wells.
$B_{\parallel }$ couples parallel and in-plane motions producing anti-crossings
in the spectrum. However, since our approach is non-perturbative, we are able
to explore the entire frequency range. For high frequencies $\omega$, we
reproduce the well known results of perfect DL given by zeroes of a Bessel
function. We find also that the system exhibits DL at the same values of the
field $F$, \emph{even as} $B_{\parallel }\neq 0 $, suggesting a hidden
dynamical symmetry in the system which we identify with different parity
operations. Symmetries under general parity operations explain many of the
features in the spectra, and their overall behavior under magnetic field. The
return times for the electron at various values of field exhibit interesting
and complex behavior which is also studied in detail. We find that smaller
$\omega $ shifts the DL points to lower $eFd/\hbar \omega $ ratios, and more
importantly, yields poorer (less effective) localization by the field, while
other states change also physical character. We analyze the explicit time
evolution of the system, monitoring the elapsed time to return to a given well
for each Landau level, and find non-monotonic behavior for decreasing
frequencies.
\end{abstract}


\maketitle

\section{Introduction}

The dynamics of charged particles in semiconductor quantum well structures
subject to a time-dependent external electric field has been a subject of
intense research.\cite{1,2,3,4,5,6,7,8,9,9a,9b,9c,10} The progress of
techniques in nanoscale lithography and the development of the free-electron
lasers (FEL)\cite{1} which can be continuously tuned in the terahertz (THz)
range, have made possible the systematic study of effects only present in
intense alternating fields in this domain. Examples of these effects are the
coherent suppression of tunneling (or \emph{dynamic localization}) despite
interwell tunneling in the structure,\cite{11,11a,11b,12} the collapse of
minibands in superlattices,\cite{13} absolute negative conductance,\cite%
{3,13a} and photon-assisted-tunneling (PAT) in resonant tunneling diodes, %
\cite{14} the AC Stark Effect,\cite{19a} and many others. These effects have
been predicted and partially experimentally verified.\cite{15,16,17,18,19}

Intense AC fields on semiconductor structures have been also studied in
connection with photon-assisted tunneling experiments with magnetic field
applied either parallel ($B_{\parallel }$),\cite{20} or perpendicular ($%
B_{\perp }$),\cite{21} to the interfaces. Quantum wells in tilted magnetic
fields are also of great current interest\cite{22,23,24,25} as experimental
probes of the transition to quantum chaos in a mesoscopic system, in the
presence of a static electric field. In this work, we study the combined
influence of tilted magnetic field and strong AC fields and find quite a
rich behavior for different parameters in the problem.

A double quantum well (DQW) can be treated to lowest approximation by
considering only the first level in each well, resulting in a two level
system with a splitting $\Delta $ due to the inter-well tunneling. This
two-level dynamics, moreover, plays an important role in understanding the
behavior of more elaborate dynamical systems, which explains why this
subject has been so intensely studied (for a comprehensive review see Ref.\ %
\onlinecite{10}).

In this paper we address the problem of a double quantum well in simultaneous
tilted magnetic field and an intense AC drive in a \emph{\ non-perturbative}
approach. To do so we make use of the Floquet-Fourier formalism developed by
Shirley.\cite{26} For strong oscillating fields, the approach based on the
Floquet theory,\cite{10,26,27} has proved most useful in treating the dynamics
of such systems, although some analytical solutions have been
proposed.\cite{6,28} Shirley's approach is very convenient since a time
dependent problem can be mapped onto a time-independent infinite matrix
eigenvalue problem (eventually truncated at the desired accuracy), which yields
a series of quasienergies describing the time evolution of the system. This
formalism is also useful as the Floquet states possess either odd or even
dynamical symmetry. As a result, to explore the effects of magnetic fields, we
use a generalized parity operator and its eigenfunctions to provide
understanding about crossings in the quasienergy spectra. We are specially
interested in the dynamics of the system in the presence of the intense AC
electric field and how it changes with the applied tilted magnetic field.
Unlike most of the previous studies of AC-driven systems, where the main
attention was paid to the high frequency regime, our approach includes the full
range of frequency values (within the single quantum well level approximation).

We begin in section \ref{Theory} with a complete description of the model
used, discussing some important properties like the parity of the system. We
find in fact that this system can be understood using two kinds of parity
operators, one when there is no parallel magnetic field applied ($%
S_{p}^{\prime }$) and a full parity operator in the other case ($S_{p}$).
Based in the two level approximation, we also develop an analytical
description that yields the degeneracy in the quasienergy for either the
high frequency ($\Delta /\hbar \omega \ll 1$) or weak electric field ($%
eFd/\hbar \omega \ll 1$) regimes, and it is in complete agreement with our
numerical calculations.

In section \ref{results} we make a systematic study of the system exploring
both the quasienergy spectra and the explicit time evolution. We find that
in the absence of a parallel magnetic field, as the Landau levels are
conserved, the problem can be understood in terms of two level systems for
each Landau level. When the parallel magnetic field is turned on, however,
the crossing levels which originated from the Landau level conservation (for
$B_{\parallel }=0$), start mixing and develop anticrossings. Thus, the
symmetry changes from $S_{p}^{\prime }\rightarrow S_{p}$ due to $%
B_{\parallel }$, lending a transition from crossing to anticrossing of
levels as the AC field amplitude is varied. For all fields, however, we find
a number of crossings, normally associated to dynamic localization, which
show no crossing-anticrossing transition and little change in dynamics when $%
B_{\parallel }$ is turned on. We have also performed a complete study of the
change in the field condition for dynamic localization, as function of the
ratio $\Delta /\hbar \omega $. We find a monotonic drop in the field values
needed to achieve localization, as the ratio $\Delta /\hbar \omega $
increases (smaller frequency), with successive low field DL points
disappearing as $\omega $ decreases. We are able to provide an empirical
equation that fits surprisingly well the numerical results, although an
analytical description valid in the different limits is also presented. An
analysis of the `quality' of DL for various AC fields is provided by a study
of the minima in the probability of a particle initially in the left well to
remain there. This result shows a surprising and non-understood behavior. As
expected for the high frequency limit, the dynamic localization is very well
defined (fully localizing the particle in a well, and at fields given by the
zeros of the Bessel function).\cite{10} However, for low frequency, the
dynamical localization becomes poorer, although the particle never fully
leaves the well, and presents small `revivals' for the first and third
dynamic localization points. This behavior is not intuitive and not reported
previously. The full range of complex behavior is however experimentally
relevant, as we discuss in section \ref{Conclusion}.

\section{\label{Theory}Theory}

The Hamiltonian for an electron in a double quantum well structure under a
tilted magnetic field and a strong AC field can be written as
\begin{equation}
H=H_{0}+eFz\cos (\omega t),  \label{t1}
\end{equation}%
where $H_{0}=(\mathbf{P}-e\mathbf{A})^{2}/2m^{\ast }+V_{0}$ is the
Hamiltonian for the non-driven system. We have chosen the growth direction
along the $z$-axis and the direction of the magnetic field in the $x$-$z$ plane $%
\mathbf{B}=(B_{\parallel },0,B_{\perp })$. Using the gauge $\mathbf{A}%
=(0,B_{\perp }x-B_{\parallel }z,0)$, and making use of the translational
symmetry in the $y$-direction, the Hamiltonian for the non-driven system can be
written as
\begin{eqnarray}
H_{0} &=&\frac{P_{x^{\prime}}^{2}}{2m^{\ast }}+\frac{m^{\ast }\omega _{\perp }^{2}%
}{2}x^{\prime 2}+\frac{P_{z}^{2}}{2m^{\ast }}+V_{0}\   \nonumber \\
&&+\frac{e^{2}B_{\parallel }^{2}}{2m^{\ast }}z^{2}-\frac{e^{2}B_{\parallel
}B_{\perp }}{m^{\ast }}zx^{\prime },  \label{t2}
\end{eqnarray}%
where $\omega _{\perp }=eB_{\perp }/m^{\ast }$ and $x^{\prime }=x+\hbar
k_{y}/eB_{\perp }$, with $k_{y}$ conserved throughout. In the absence of $%
B_{\parallel }$, and allowing for only one state in each well of the DQW,
Eq.\ (\ref{t2}) is exactly soluble and its eigenenergies are $\hbar \omega
_{\perp }(n+1/2)\pm \Delta /2$ (harmonic oscillator in plane with the
symmetrical and antisymmetrical solutions of the DQW), where $\Delta $ is
the splitting due to the tunneling. Inclusion of additional levels in each
quantum well is straightforward, and it does not add to the qualitative
behavior and conclusions described here (except for possible multi-photon
resonances at high field, the AC Stark effect\cite{19a}). Moreover, thin
wells will likely have others levels at much higher energies, and our
treatment here will be quantitatively accurate.

The dynamics of the system is governed by the time-dependent Schr\"{o}dinger
equation
\begin{equation}
i\hbar \frac{\partial |\psi \rangle }{\partial t}=H|\psi \rangle .
\label{t5}
\end{equation}%
Since $H$ is periodic in time ($H(t)=H(t+\tau )$, where $\tau =2\pi /\omega $
is the period) we can make use of the standard Floquet theory\cite{11,26,27}
and write the eigenfunction $|\psi \rangle $ as
\begin{equation}
|\psi \rangle =\exp (-i\varepsilon t/\hbar )|u\rangle ,  \label{t6}
\end{equation}%
which allows us to rewrite Eq.\ (\ref{t5}) as
\begin{equation}
(H-i\hbar \partial _{t})|u\rangle =\varepsilon |u\rangle ,  \label{t5a}
\end{equation}%
where $|u(t)\rangle =|u(t+\tau )\rangle $ is also periodic in time, with the
same period $\tau $, and $\varepsilon $ is a real-valued parameter termed
the Floquet characteristic exponent, or the quasienergy.\cite{10} The term
quasienergy reflects the formal analogy with the quasi-momentum $k$,
characterizing Bloch eigenstates in a periodic solid.\cite{29} Following
this analogy we can see that $|u_{(m)}\rangle =\exp (im\omega t)|u\rangle $
is also solution of Eq.\ (\ref{t5a}), with eigenvalues $\varepsilon
_{m}=\varepsilon +m\hbar \omega $, if $m$ is an integer number $m=0,\pm
1,\pm 2,...$. Then, by a subtraction of a suitable integral multiple of $%
\hbar \omega ,$ the quasienergy $\varepsilon $ can be mapped into a first
`Brillouin zone' obeying: $\hbar \omega /2\leq \varepsilon <\hbar \omega /2$%
. For vanishing intensity of the AC field $F,$ the Floquet states are
connected with the stationary states $|\alpha ^{(0)}\rangle $ of $H_{0}$ by $%
|u_{(l)}\rangle =\exp (il\omega t)|\alpha ^{(0)}\rangle $, and the
quasienergies with the unperturbed states by $\varepsilon _{l}=E_{\alpha
}^{0}+l\hbar \omega ,$ where the index $l$ counts how many AC energy quanta
have to be subtracted from the unperturbed energy $E_{\alpha }^{0}$ in order
to arrive at the first Brillouin zone (the `photon index').\cite{10}

Following the procedure developed by Shirley,\cite{26} we solve Eq.\ (\ref%
{t5a}) by expanding the function $|u\rangle $ in a Fourier series for the
temporal component, and linear combinations of the $B_{\parallel }=0$
solutions of $H_{0}$ for the spatial component,
\begin{equation}
|u\rangle =\sum_{s,n,m}C_{s,n,m}|s,n,m\rangle ,  \label{t7}
\end{equation}%
where $|s\rangle $ refers to the symmetrical $(|S\rangle )$ and
antisymmetrical $(|A\rangle )$ eigenfunctions of the DQW, $|n\rangle $ is
the Landau level function, and $|m\rangle $ is a Fourier expansion `state' ($%
\langle t|m\rangle =\exp (im\omega t)$). This leads to the time-independent
infinite matrix eigenvalue equation
\begin{eqnarray}
&&\sum_{s,n,m}\left\{ \left[ E_{s}+\hbar \omega _{\perp }(n+\frac{1}{2}%
)-\varepsilon +m\hbar \omega \right] \delta _{s,s^{\prime }}\delta
_{n,n^{\prime }}\delta _{m,m^{\prime }}\right.  \nonumber \\
&&+\left[ \frac{e^{2}B_{\parallel }^{2}}{2m^{\ast }}z_{s,s^{\prime
}}^{2}\delta _{n,n^{\prime }}-\frac{e^{2}B_{\parallel }B_{\perp }}{m^{\ast }}%
z_{s,s^{\prime }}x_{n,n^{\prime }}^{\prime }\right] \delta _{m,m^{\prime }}
\nonumber \\
&&+\left. \frac{eFz_{s,s^{\prime }}}{2}\left( \delta _{m,m^{\prime
}-1}+\delta _{m,m^{\prime }+1}\right) \delta _{n,n^{\prime }}\right\}
C_{s,n,m}=0,  \label{t8}
\end{eqnarray}%
where $z_{s,s^{\prime }}^{2}=\langle s|z^{2}|s^{\prime }\rangle ,$ $%
z_{s,s^{\prime }}=\langle s|z|s^{\prime }\rangle $, $x_{n,n^{\prime
}}^{\prime }=\langle n|x^{\prime }|n^{\prime }\rangle $, and $E_{s}=\pm
\Delta /2$.

\subsection{Floquet states and parity}

The quasienergies in Eq.\ (\ref{t5a}) can be regarded as the eigenvalues of
a stationary problem analogous to a time-independent Schr\"{o}dinger
equation, with $|u\rangle $ playing the role of stationary states of the
operator $\mathcal{H}=H-i\hbar \partial _{t}$. As a consequence of $[%
\mathcal{H},S_{p}]=0,$ where $S_{p}$ is the parity operator defined by $%
S_{p}:(x^{\prime }\rightarrow -x^{\prime },z\rightarrow -z,t\rightarrow
t+\tau /2)$, the Floquet states $|u\rangle $ have even or odd parity under $%
S_{p}$.\cite{11,27,30,31} Since our expansion basis (Eq.\ (\ref{t7})) has
also a definite parity under $S_{p}$, the matrix (\ref{t8}) can be separated
into two blocks for subspaces of different symmetries, one for $s+n+m=even$
and another for $s+n+m=odd$. This result is very useful when interpreting
the quasienergy spectrum, which contains multiple crossings (that come from
different symmetries), as well as avoided crossings associated with mixing
between same-parity states.

It is important to note also that for vanishing $B_{\parallel }$, the
parallel and perpendicular motions are completely decoupled, and in this
case the problem reduces to solving a DQW system under intense AC field for
each in-plane Landau level separately. In this $B_{\perp }$-only case, the
parity operator can be defined as $S_{p}^{\prime }:(z\rightarrow
-z,t\rightarrow t+\tau /2)$, and as $[\mathcal{H},S_{p}^{\prime }]=0$, this
yields separate parities under the condition $s+m=even$ or $s+m=odd$.

\subsection{Evolution operator and degeneracies}

In the Floquet representation, the time-evolution operator $U(t,t_{0}),$
defined by $|\psi (t)\rangle =U(t,t_{0})|\psi (t_{0})\rangle ,$ with $%
U(t_{0},t_{0})=I$, where $I$ denotes the identity operator, can be expressed
as
\begin{equation}
U(t,t_{0})=\sum_{i}|u_{i}(t)\rangle \langle u_{i}(t_{0})|\exp (-i\varepsilon
_{i}(t-t_{0})/\hbar ),  \label{t9}
\end{equation}%
as it is easily checked with the help of Eq.\ (\ref{t5a}). A full period $%
U(\tau ,0)$ operator is all we need to construct a discrete quantum map,
propagating an initial state over multiples of the fundamental period.\cite%
{10} The evolution operator formulation allows one to explore the condition
of quasi-energy degeneracy, which we will see is an important necessary
condition for dynamic localization (although not sufficient).

As mentioned above, without the in-plane magnetic field, the problem reduces to
a two-level system for each Landau level driven by an external time-dependent
electric field, which has been extensively studied in the literature (for a
review, see Ref.\ \onlinecite{10}). Although this system looks simple, the
physics in it is rich, and a complete analytical solution is either lacking or
too complicated to use in the different parameter regimes.\cite{32} An
analytical extension to the solution in the high frequency limit was recently
found,\cite{33} providing a perturbative connection to this case. Here we give
an analytical solution applicable in both the high frequency and weak field
limits. For simplicity, let us consider the system in the absence of a magnetic
field, so that the Hamiltonian can be written in a two level basis as
\begin{equation}
H=H_{1}+H_{2}=\frac{\Delta }{2}\sigma _{z}+\frac{eFd}{2}\sigma _{x}\cos
(\omega t),  \label{t10}
\end{equation}%
where $\sigma _{z},$ $\sigma _{x}$ are the usual Pauli matrices.

The time evolution operator $U$ satisfies the Schr\"{o}dinger equation in
the operator form
\begin{equation}
i\hbar \frac{dU}{dt}=HU.  \label{t11}
\end{equation}%
The difficulty in solving the problem lies in the non-commutation between $%
H_{1}$ and $H_{2}$. For only $H_{1}$ or $H_{2}$, the solution is $U_{1}=\exp
(-i\Delta \sigma _{z}t/2\hbar )$, or $U_{2}=\exp (-i\int_{0}^{t}eFd\cos
(\omega t)\sigma _{x}dt/2\hbar )$, respectively. We write the time evolution
operator in the form $U=U_{0}U^{\prime }=U_{1}U_{2}U^{\prime }$, with $%
U^{\prime }$ satisfying the equation
\begin{equation}
i\hbar \frac{dU^{\prime }}{dt}=H_{\mathrm{eff}}U^{\prime },  \label{t12}
\end{equation}%
where $H_{\mathrm{eff}}=U_{0}^{-1}H_{2}U_{0}-H_{2}$. In the high frequency
limit ($\Delta /\hbar \omega \ll 1$) or weak electric field regime ($%
eFd/\hbar \omega \ll 1$), we can write
\begin{equation}
U^{\prime }=\exp \left( -\frac{i}{\hbar }\int_{0}^{t}H_{\mathrm{eff}%
}(t^{\prime })dt^{\prime }\right) .  \label{t13}
\end{equation}%
By diagonalizing $U(\tau )$, we can obtain the quasienergies. After some
tedious but straightforward algebra, we find that the condition for
degeneracy of quasienergy levels, ($\Delta _{\mathrm{eff}}/\hbar \omega =0$ (%
$\mathrm{mod}(2\pi )$) is given by the condition
\begin{eqnarray}
&&\frac{\Delta }{\hbar \omega }\int_{0}^{2\pi }\cos \left( \frac{\Delta }{%
\hbar \omega }t^{\prime }\right) \cos \left( \frac{eFd}{\hbar \omega }\sin
(t^{\prime })\right) dt^{\prime }  \nonumber \\
&&+\frac{\Delta }{\hbar \omega }2\pi -\sin \left( \frac{\Delta }{\hbar
\omega }2\pi \right) =0\text{\ }(\mathrm{mod}(2\pi )).  \label{t14}
\end{eqnarray}%
Since the energy splitting $\Delta _{\mathrm{eff}}$ is due to tunneling and
level mixing by the AC field, this quasienergy degeneracy means a suppression
of tunneling, or in other words, this is a condition for the dynamic
localization to occur. It is easy to see that in the high frequency limit, Eq.\
(\ref{t14}) yields the well-known condition $J_{0}(eFd/\hbar \omega )=0$. In
the weak field limit, one obtains the result $\Delta /\hbar \omega =n$, $n$
being an even integer number (for $n$ odd there is only solution for $F=0$ and
does not lead to dynamic localization). As we will see later, these limiting
degeneracy conditions are in fact realized in our numerical work, and allow one
to understand the physics of dynamic localization.

\section{\label{results}Results and discussions}

In order to perform numerical calculations, we choose a symmetric double
quantum well ($50-40-50$\ \AA\ GaAs/Al$_{0.3}$Ga$_{0.7}$As/GaAs) with effective
electron mass $m^{\ast }=0.067\ m_{0}$ ($m_{0}$ is the electron mass),
$d=2\langle S|z|A\rangle \simeq 90$\ \AA\ is the mean separation
between the left and right wells, and $\Delta =8.87$\ meV.\footnote{%
This value is obtained from a direct solution of such two-well potential.
The next excited states are in the continuum, out of the potential well and
basically uncoupled in the range of fields and frequencies we consider here.}
We have performed a systematic study varying the external parameters ($%
B_{\parallel }$, $B_{\perp }$, $\omega $ and $F$) and analyze the results
for different ranges of the ratio $\Delta /\hbar \omega $. It is important
to emphasize that our numerical model is valid for all values of the ratio $%
\Delta /\hbar \omega $. However, as will become clear in the discussion
below, the behavior of the system is markedly different for $\Delta /\hbar
\omega \lesssim 1$ and for $\Delta /\hbar \omega >1$. We first discuss the
first regime, $\Delta /\hbar \omega \lesssim 1$, as it is perhaps more
intuitive, to later look at lower frequencies.

\subsection{High frequency, $\Delta /\hbar \protect\omega <1$}

\begin{figure*}[tbp]
\includegraphics*[width=8.6cm]{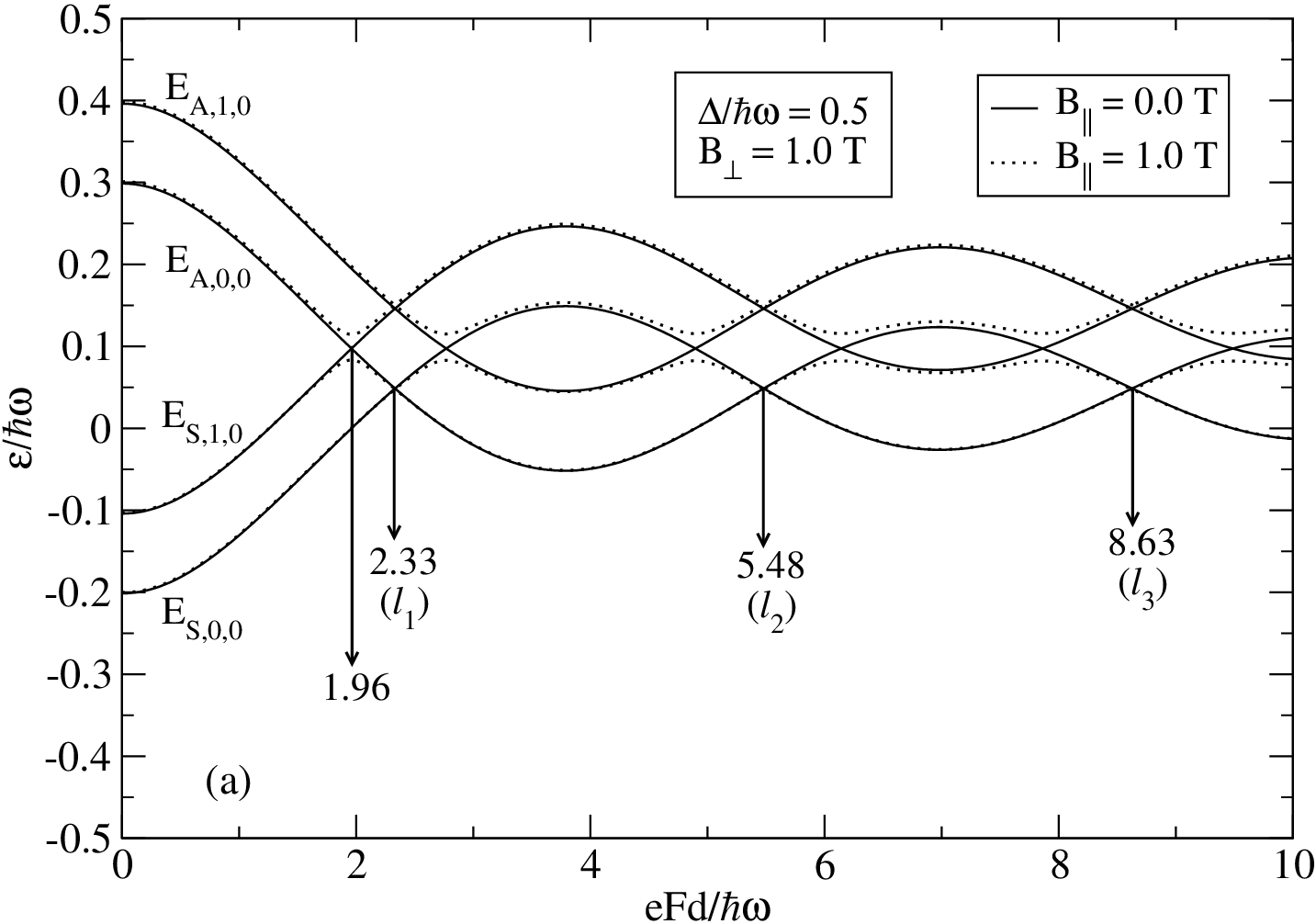}\hspace{0.1cm}
\includegraphics*[width=8.6cm]{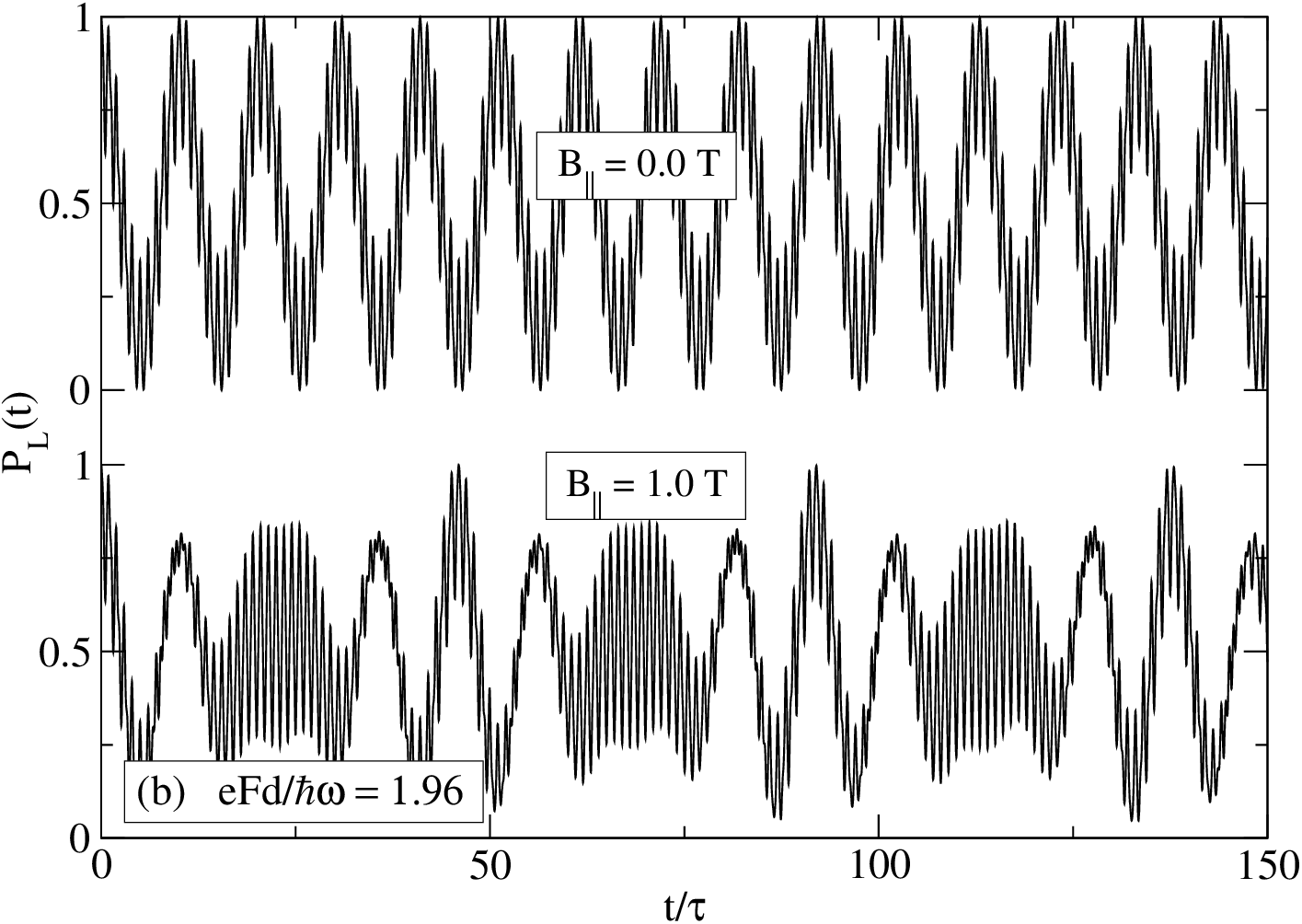}\vspace{0.15cm}
\includegraphics*[width=8.6cm]{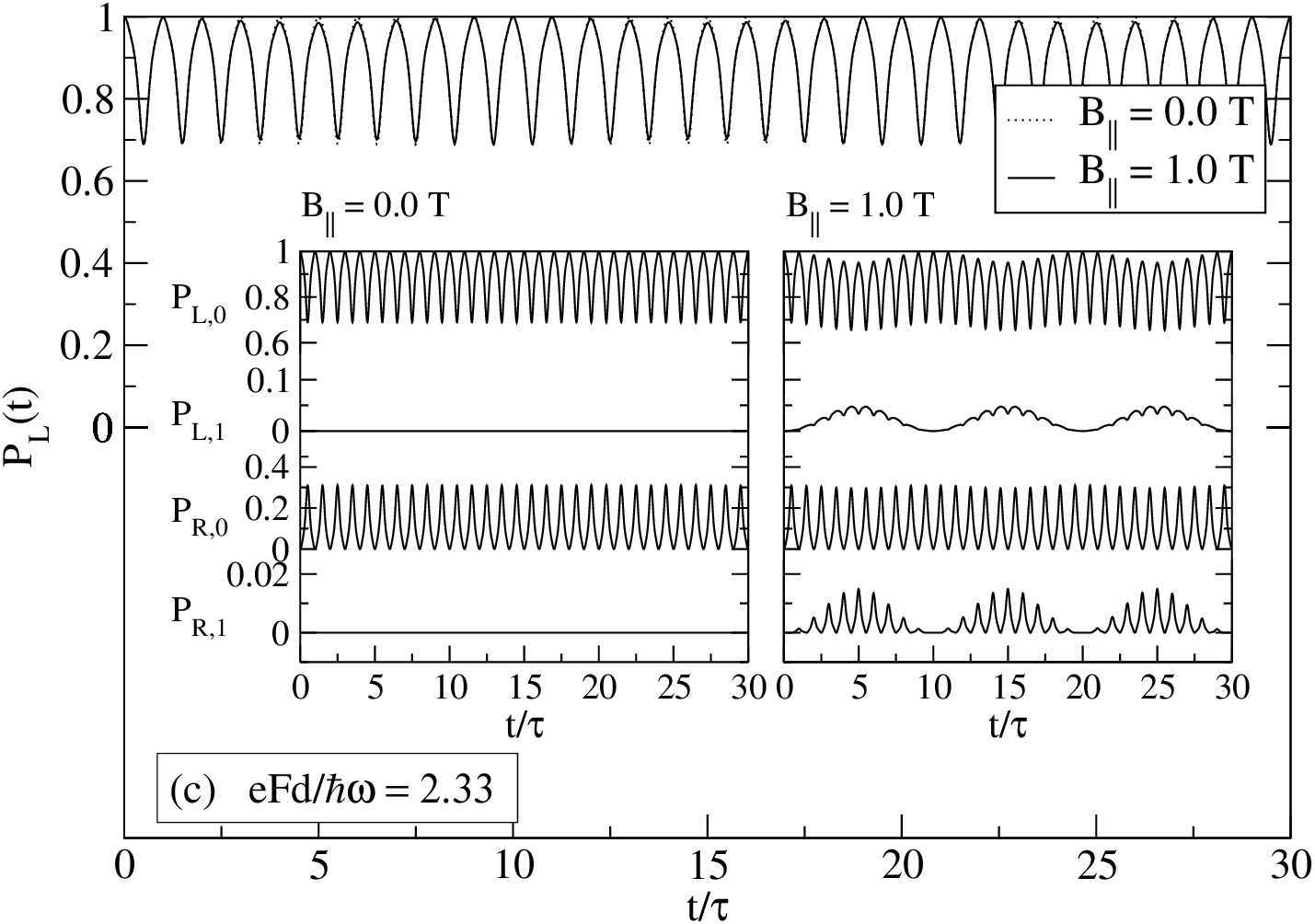}\hspace{0.1cm}
\includegraphics*[width=8.6cm]{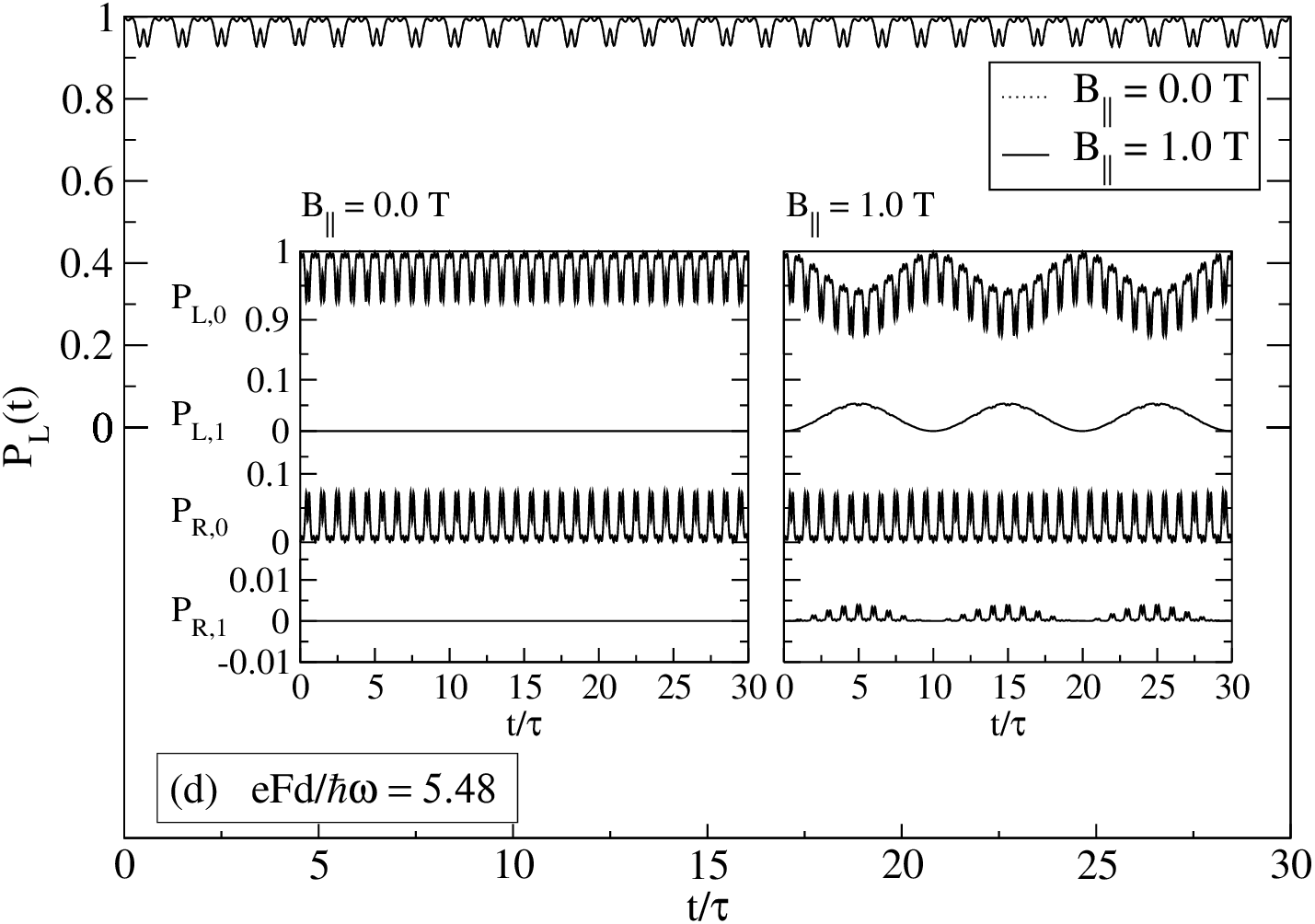}
\caption{(a) First Brillouin zone of quasienergies (in units of $\hbar
\protect\omega $) as function of $eFd/\hbar \protect\omega $ (intensity of
AC field), for $\Delta /\hbar \protect\omega =0.5$ and $B_{\perp }=1.0$~T.
The label $E_{s,n,m}$ refers to the eigenvalues of $H_{0}$ for vanishing $F$
and $B_{\parallel }$ ($s$ indicates the symmetrical or antisymmetrical
state; $n$, the Landau level index; and $m$ counts how many photon quanta
have to be subtracted or added to the unperturbed energy to arrive at the
first Brillouin zone). The main difference between $B_{\parallel }=0$ (full
line) and $B_{\parallel }=1.0$~T (dotted line) are the anticrossings in
points such as $eFd/\hbar \protect\omega \simeq 1.96$. The points $l_{i}$
indicate the dynamical localization points, and exhibit no change with $%
B_{\parallel }$. (b) Time evolution of the probability $P_{L}(t)$ of an
electron prepared in the first Landau level in the left well to be found in
the same well for $eFd/\hbar \protect\omega \simeq 1.96$ (time in units of
the AC field period $\protect\tau =2\protect\pi /\protect\omega $). (c) Same
as (b), but for $eFd/\hbar \protect\omega \simeq 2.33$, the $l_{1}$ point.
This shows dynamic localization, since the system can be found with at least
$70\%$ probability in the left well for all time. The inset figures show the
time evolution probability for all states of the system, for different
values of $B_{\parallel }$. Note that for $B_{\parallel }=1.0$~T there is a
small probability for the system to be found in the first Landau level, $%
P_{L,1}$ and $P_{R,1}$, but $P_{L}$ is left unchanged. (d) Same as (c) for $%
eFd/\hbar \protect\omega \simeq 5.48$ $(l_{2})$. The dynamic localization is
much better defined for this point.}
\label{fig1}
\end{figure*}

We start our analysis by considering the quasienergy spectrum for two Landau
levels and varying $B_{\parallel }$. In Fig.\ \ref{fig1}(a) we show the
first Brillouin zone of quasienergy in units of photon energy, for $\Delta
/\hbar \omega =0.5$ and $B_{\perp }=1.0$\ T. The labels $E_{s,n,m}$ for each
level, correspond to the \textit{z-}symmetry (S or A), the Landau level
index $n$, and the `photon index' $m$, all$\ $at $F=0$. For $B_{\parallel
}=0 $ the parallel and perpendicular motions are completely decoupled and
the Landau level number is conserved (only transitions with the same Landau
level index are allowed). This explains the level crossings at points such
as $eFd/\hbar \omega \simeq 1.96$ and all others not explicitly labeled,
according to $S_{p}^{\prime }$ symmetry (that is, levels crossing have
different symmetry under $S_{p}^{\prime }$). When we turn on the parallel
magnetic field, however, the parallel and in-plane motions are coupled, so
that the Landau level index is no longer a good quantum number. This
produces anticrossings in the quasienergy spectrum at points such as $%
eFd/\hbar \omega \simeq 1.96$, while still keeping crossings at the points
denoted by $l_{i}$ (localization points), according to their $S_{p}$
symmetry. Here we see explicitly that $B_{\parallel }$ produces a change of
symmetry $S_{p}^{\prime }\rightarrow S_{p}$. Thus the system for $eFd/\hbar
\omega \simeq 1.96$ shows a transition from level crossing (the states are
odd and even in $S_{p}^{\prime }$) to level anticrossing (the states are
both odd under $S_{p}$) due to $B_{\parallel }$. It is easy to see that
there is no such transition for $l_{i}$ points (dynamic localization
points), since at $l_{i}$ states have opposite parity under both $%
S_{p}^{\prime }$ and $S_{p}$ symmetry operations. This is the essential
difference between the point $eFd/\hbar \omega \simeq 1.96$ and $l_{i}$.

In order to better understand the physical behavior of the system at the
various crossings, we calculate the time evolution of the system for some
particular points (points indicated by arrows in Fig.\ \ref{fig1}(a)), using
Eqs.\ (\ref{t6}) and (\ref{t7}). Here, the system is prepared in the first
Landau level of the left well at $t_{0}$, $|\psi (t_{0})\rangle =|L,0\rangle
,$ where $|L,0\rangle =|L\rangle \otimes |0\rangle $ with $|L\rangle =1/%
\sqrt{2}(|S\rangle -|A\rangle )$, and $|0\rangle $ refers to the lowest Landau
level for the in-plane motion. We investigate how the probability of finding
the particle in some state $|\beta \rangle =|l,n\rangle $ $(l=L,R$ and
$n=0,1,2,...)$, $P_{\beta }(t)=|\langle \beta |\psi (t)\rangle |^{2}$, changes
with time. We also investigate the \emph{total} probability of the system to
remain in the left well
\begin{equation}
P_{L}(t)=\sum_{n}|\langle L,n|\psi (t)\rangle |^{2}.  \label{r1}
\end{equation}%
The probability for staying in the left well at the crossing/anticrossing point
$eFd/\hbar \omega \simeq 1.96$ can be seen in Fig.\ \ref{fig1}(b). We note a
large change in its behavior with $B_{\parallel }$. For $B_{\parallel }=0$, the
time evolution shows that the system can be found in both left and right wells
with the same probability and with a rather simple periodic behavior. The
particle is effectively tunneling back and forth with a period $\approx
10\tau =20\pi /\omega $, and with fast oscillations of period $\tau $. For $%
B_{\parallel }\neq 0$ the probability presents a more complex behavior, with
quasiperiodic oscillations. Notice in Fig.\ \ref{fig1}(b) that the system
does not have short-term periods, and a long-time plot of $P_{L}(t)$ in this
case yields periods $\approx 1000\tau $ and quasi-periodic motion at $%
\approx 46\tau $. We should also point out that the detailed frequency
content will become even richer if one includes higher Landau levels
(ignored here for simplicity).

In Fig.\ \ref{fig1}(c) we show results for the point $eFd/\hbar \omega \simeq
2.33$, the first localization point $l_{1}$. We can clearly see that in this
case, the probability for the system to stay in the left well is very high and
never goes to zero, indicating a dynamic localization of the charge, ensued by
the AC field despite the interwell tunneling allowed in the structure. It is
important to note that the parallel magnetic field has no effect on the
position of this dynamic localization point. However, as can be seen in the
right inset, there is a small probability for the system to be found in the
second Landau level, $n=1$. At this point, the system
exhibits small transitions between Landau levels, but in the \emph{same well}%
, so that the localization is basically invariant under the presence of the
parallel magnetic field (although its return period to the $n=0$ level is
now longer, as seen in the right inset in \ref{fig1}(c)).

For the crossing point at $eFd/\hbar \omega \simeq 5.48$ $(l_{2})$, Fig.\ %
\ref{fig1}(d) shows similar results, except for the fact that the time
evolution presents better dynamic localization (notice minima in $%
P_{L}\gtrsim 0.92$). We in fact find that in general, dynamic localization
points are `better localized' for higher $F$ values, resulting in higher minima
for $P_{L}$ and/or weaker $P_{R}$ maxima, although there is a few exceptional
examples, as we will see in Fig.\ \ref{fig7}.\footnote{We should point out that
our numerical finding routine of the crossing points has finite accuracy, and
the shallower slopes in the spectrum allow a finer, more accurate,
determination of the degeneracy point, which may result in slightly better
defined localization at higher $F$. Further refinements of the $l_{i}$ points
yields generally better localization, although this improvement saturates
especially for large $\Delta /\hbar \omega $ ratios (see discussion below for
Fig.\ \ref{fig7}), and we are certain the behavior is physical and not a
numerical artifact.}

\begin{figure}[tbp]
\includegraphics*[width=8.6cm]{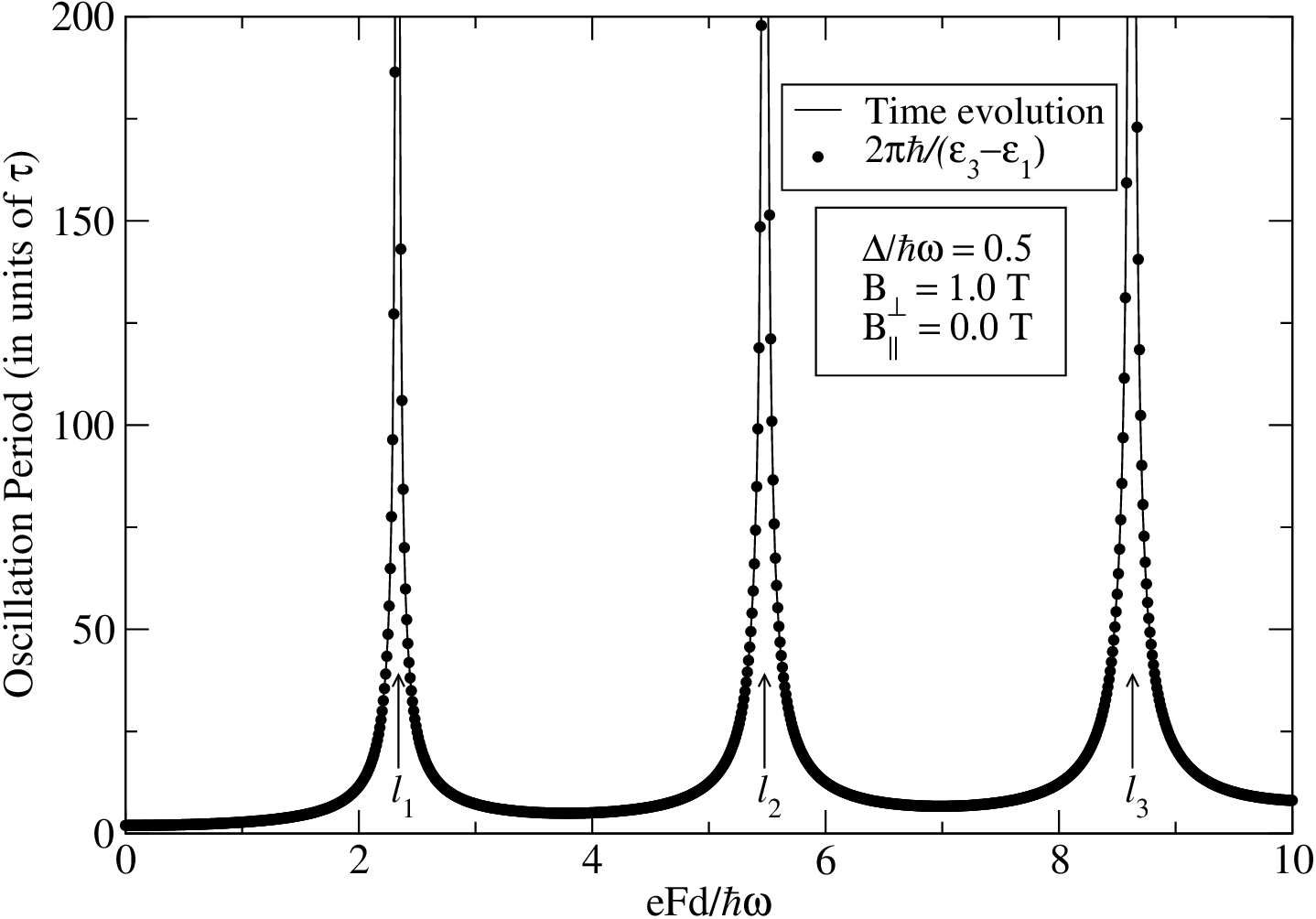}
\caption{Oscillation period as function of $eFd/\hbar \protect\omega $
(intensity of AC field), for $\Delta /\hbar \protect\omega =0.5$, $B_{\perp
}=1.0$~T and $B_{\parallel }=0$. At the points $l_{i}$ the oscillation
period tends to infinity, showing the dynamic localization. Solids circles
are the results from Eq.\ (\ref{r2}), while the full line is obtained from
the direct time evolution.}
\label{fig2}
\end{figure}

Another approach to monitor the dynamic localization is to look for the
oscillation period in $P_{L}$, defined naturally as the time for the
electron with initial state in the left well to go to the right well (with
amplitude of 100\%) and \emph{fully} back (with amplitude of 100\%). In
Fig.\ \ref{fig1}(b), for $B_{\parallel }=0$, it is easy to see that the
oscillation period is about $10\tau $ as mentioned before. Figure \ref{fig2}
displays the return periods obtained from direct analysis of $P_{L}$ for
different values of the AC field amplitude $F$, for $\Delta /\hbar \omega
=0.5$. Notice that the generic period is rather short, except for values of
the field which produce dynamic localization $l_{i}$. Near these values, the
return period in $P_{L}$ quickly increases, suggesting indeed a divergent
period right at the $l_{i}$ values, and fully consistent with dynamic
localization behavior. Note that in Fig.\ \ref{fig1}(c) and (d), for
example, we cannot see the slow envelope which yields a finite return time,
since the system never goes to the right well with more than 50\%, and all
the oscilations remaining correspond to the high frequency with period $\tau
$.

This oscillation period can also be obtained from inspection of the
quasienergy spectrum. As $B_{\parallel }=0$, for example, only transitions
with the same Landau level index are allowed, and the oscillation period $
O_{p}$ can written as
\begin{equation}
O_{p}=\frac{2\pi \hbar }{(\varepsilon _{3}-\varepsilon _{1})},  \label{r2}
\end{equation}
where $\varepsilon _{3}$ refers to the quasienergy labeled $E_{A,0,0}$ and $
\varepsilon _{1}$ refers to $E_{S,0,0}$ in Fig.\ \ref{fig1}(a)
(quasienergies that evolve from the unperturbed state with the same Landau
level index). The solid circles in Fig.\ \ref{fig2} illustrate the values
from Eq.\ (\ref{r2}), in complete agreement with the direct numerical
calculation. We emphasize that as these results are rather intuitive, and
expected, they provide assurance that the return period is a good direct
measure of the degree of localization. As the time evolution of $P_{L}(t)$
becomes much more complex for different $\Delta /\hbar \omega $ ratios (or $%
B_{\parallel }\neq 0$), one can rely in the physical insights it provides.

\begin{figure}[tbp]
\includegraphics*[width=8.6cm]{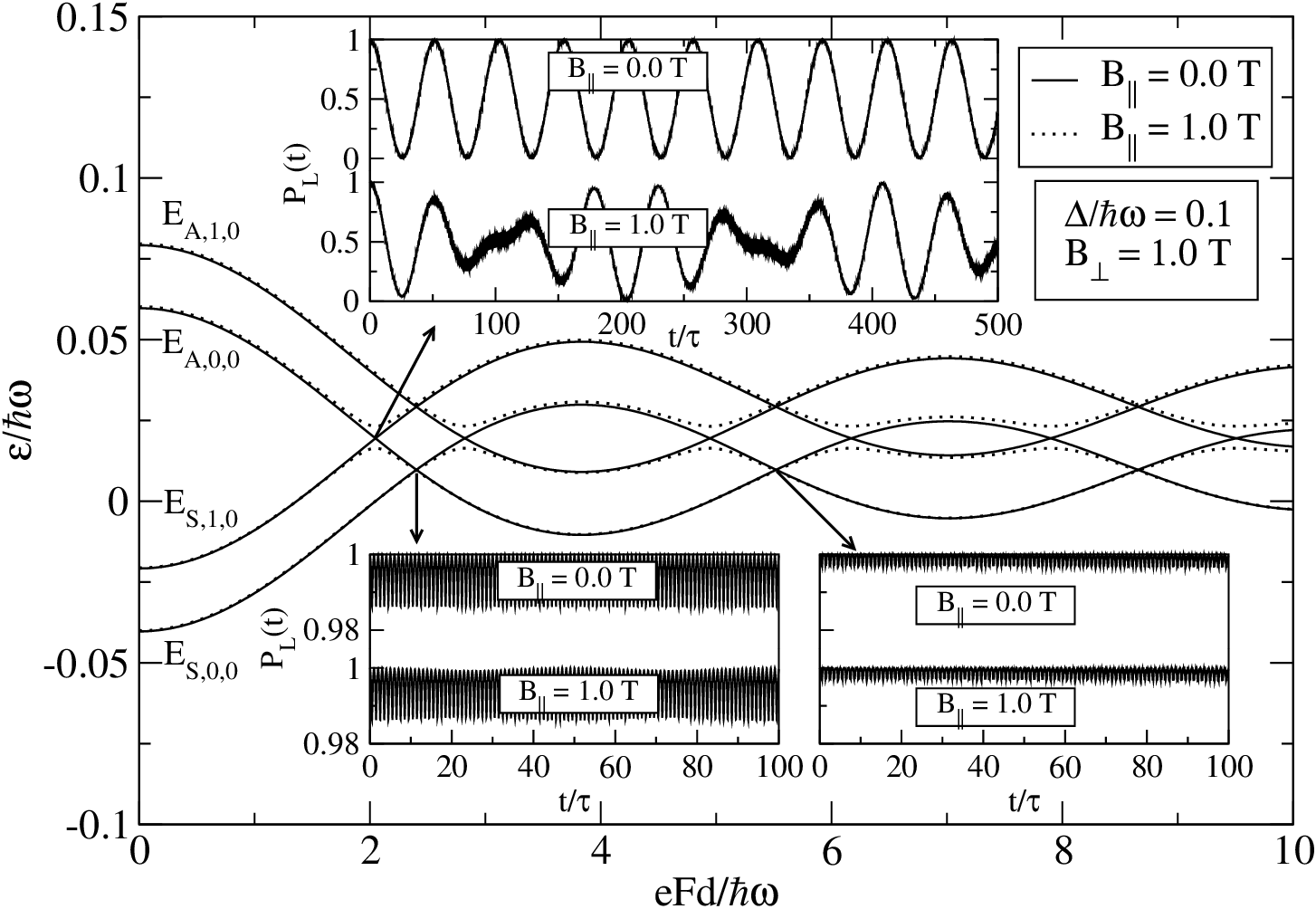}
\caption{First Brillouin zone of quasienergies (in units of $\hbar \protect%
\omega $) as function of $eFd/\hbar \protect\omega $ (intensity of AC
field), for $\Delta /\hbar \protect\omega =0.1$ and $B_{\perp }=1.0$~T.
Solid ($B_{\parallel }=0$) and dotted ($B_{\parallel }=1.0$~T) lines show
quasienergy behavior with in-plane magnetic field. The inset figures show
the time evolution $P_{L}(t)$ of the system initially in the left well for
the points indicated by arrows.}
\label{fig3}
\end{figure}

For the high frequency limit $(\hbar \omega \gg \Delta )$, as we discussed
before, the localization points tend to the zeros of the Bessel Function $%
J_{0}(eFd/\hbar \omega )$ for a double quantum well (two level problem).
This approximation is quite good, even with a tilted magnetic field, as we
would expect from the different parity of the crossing states under $S_{p}$,
and as we can see in Fig.\ \ref{fig3} for $\Delta /\hbar \omega =0.1.$ The
dynamic localization regime here is also much better than at $\Delta /\hbar
\omega =0.5$, as shown in the inset (with $P_{L\min }\gtrsim 0.98$ at $l_{1}$%
, $eFd/\hbar \omega \simeq 2.40$, for example). As we will see below, the
\emph{degree} (or `quality') of dynamic localization decreases for large $%
\Delta /\hbar \omega $, even as the quasienergy levels still cross.

\subsection{Low frequency, $\Delta /\hbar \protect\omega >1$}

\begin{figure}[tbp]
\includegraphics*[width=8.6cm]{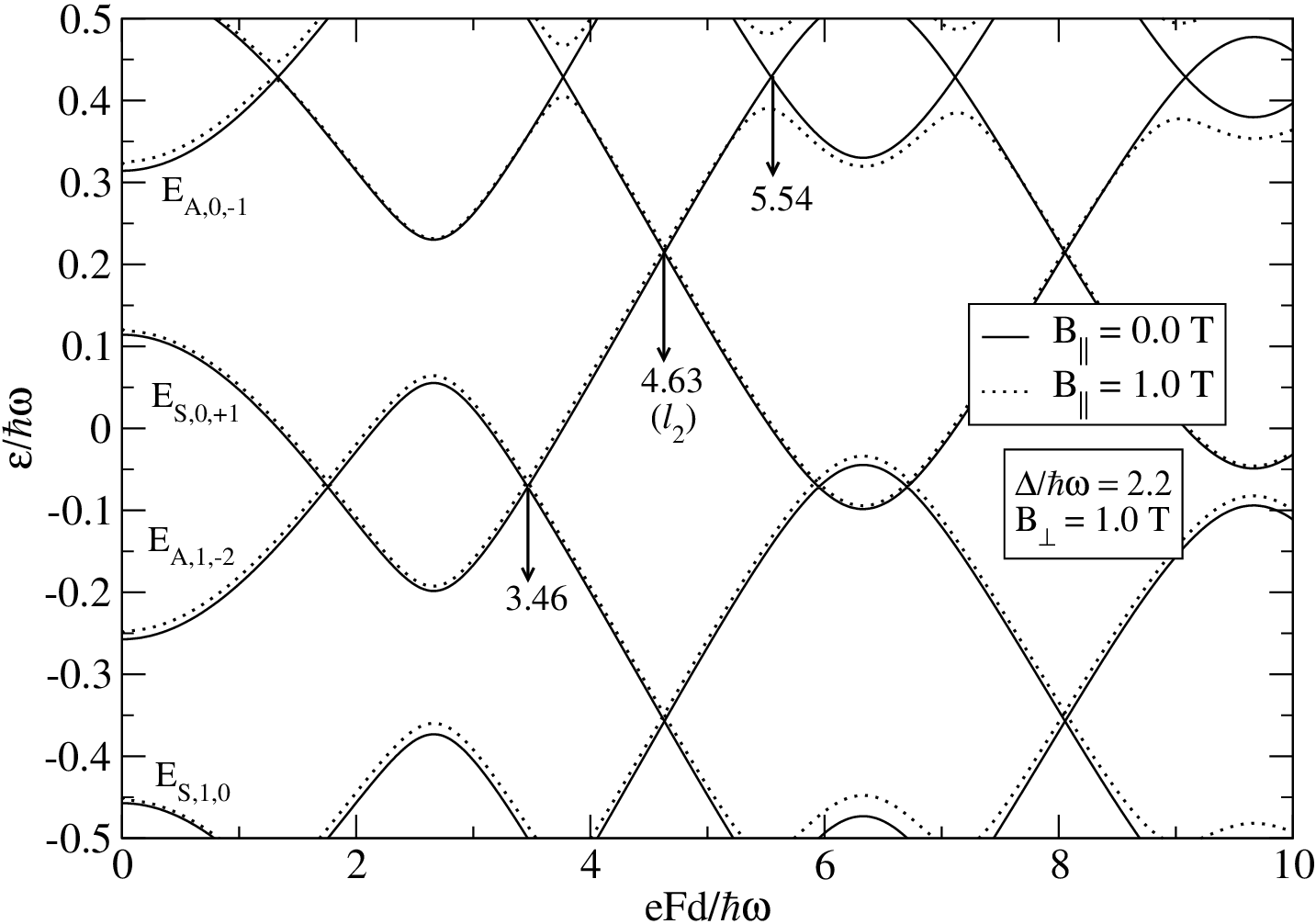}
\caption{First Brillouin zone of quasienergies (in units of $\hbar \protect%
\omega $) as function of $eFd/\hbar \protect\omega $ (intensity of AC
field), for $\Delta /\hbar \protect\omega =2.2$ and $B_{\perp }=1.0$~T. The $%
l_{2}$ label refers to a localization point and it presents no change with $
B_{\parallel }$. Others crossing/anticrossing points are also indicated (see
text). Notice labels $E_{s,n,m}$ include different photon index $m$.}
\label{fig4}
\end{figure}

In Fig.\ \ref{fig4} we show the quasienergy spectrum for $\Delta /\hbar
\omega =2.2.$ Up to now, all the spectra analyzed could have been described
without the inclusion of the photon index, since it is zero in all cases. In
the present case of large $\Delta /\hbar \omega $, however, we need to be
more careful, since the first Brillouin zone of quasienergies involves
multiple `photon replicas' and various levels could be involved in crossings
for \emph{different} photon index. To understand crossing and anticrossing
we need to explore the full symmetry of the states under the parity operator
$S_{p}$.

\begin{figure}[tbp]
\includegraphics*[width=8.6cm]{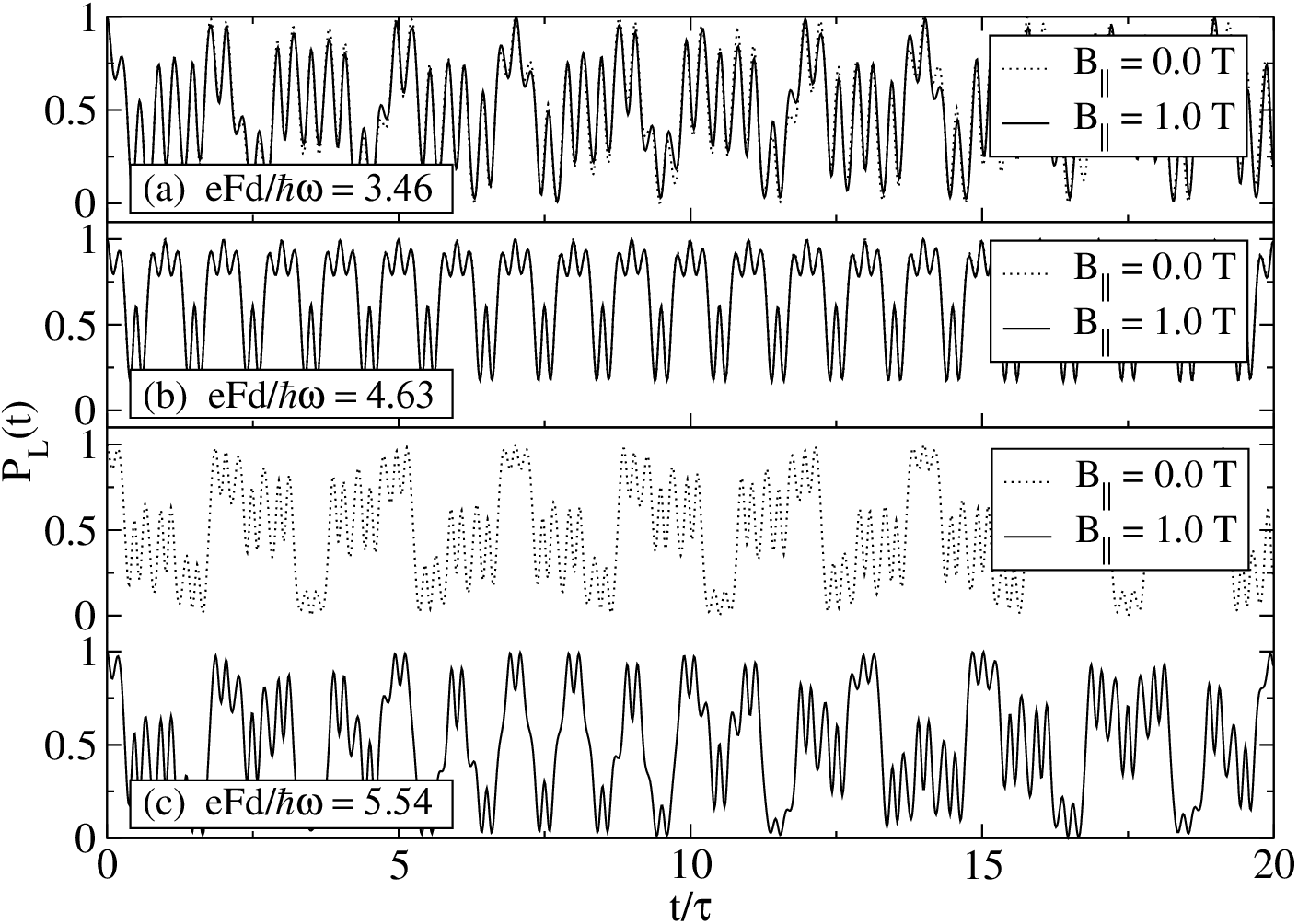}
\caption{Time evolution of the probability $P_{L}(t)$ of an electron
prepared in the first Landau level in the left well, and to be found
projected in the same well: a) for $eFd/\hbar \protect\omega \simeq 3.46$;
b) for $eFd/\hbar \protect\omega \simeq 4.63$; c) for $eFd/\hbar \protect%
\omega \simeq 5.54$. Only (b) exhibits dynamic localization, although
incomplete.}
\label{fig5}
\end{figure}

In Fig.\ \ref{fig5} we choose some quasienergy crossing points (indicated by
arrows in Fig.\ \ref{fig4}) and carry out the time evolution. For the point
at $eFd/\hbar \omega \simeq 3.46$, the time evolution shows that it is a
`normal' point, namely with the particle fully jumping to the right well, so
that the minimum $P_{L}=0$. This behavior suggests that the states crossing
have different symmetry which the tilted magnetic field seems not to affect.
On the other hand, the crossing point $eFd/\hbar \omega \simeq 4.63$ $%
(l_{2}) $ is similar to the dynamic localization points discussed before, in
the sense that the probability of staying in the left well never reaches
zero (although here the localization is somewhat poor, with minimum $%
P_{L}\simeq 0.2$). Moreover, the point $eFd/\hbar \omega \simeq 5.54$ is
also a `normal' case of crossing/anticrossing produced by the inclusion of
parallel magnetic field, so that the probability of staying in the left well
changes when we include $B_{\parallel }$, and presents complex
time-dependent behavior. For any value of $B_{\parallel }$ however, the
minimum $P_{L}\simeq 0$, and the particle fully tunnels back and forth
between the wells.

It is important to point out that the crossing of levels at $eFd/\hbar
\omega \simeq 3.46$, and similar others, is associated with their different
parity under $S_{p}$. For example, at the point $\simeq 3.46$, the levels
crossing are those with $E_{A,1,-2}$ and $E_{S,0,+1}$ at $F=0.$ These states
have then \emph{even} and \emph{odd} parity under $S_{p}$, respectively (as
given by $s+n+m$), which remains true for $F\neq 0$. On the other hand, the
crossing (anticrossing) at $\simeq 5.54,$ and others, correspond to Landau
levels conservation at $B_{\parallel }=0$ (or mixing for $B_{\parallel }\neq
0$ ), and the associated parity under $S_{p}^{\prime }$.

Another region of interest is $\Delta /\hbar \omega \simeq odd$ integer. In
that case, different photon replicas are nearly degenerate but bring
together states with the same symmetry under $S_{p}$ (or $S_{p}^{\prime }$).
The same parity allows mixing of the levels, producing an interesting
anticrossing behavior for $F\simeq 0$, as shown in Fig.\ \ref{fig6} for $%
\Delta /\hbar \omega =0.9$ and $\Delta /\hbar \omega =1.1$. There we see that
in (a), the states labeled $s=S$ $($or $A)$ have an upward (downward)
dispersion for $F\simeq 0$ (similar to all cases with $\Delta /\hbar \omega
<1,$ such as Figs.\ \ref{fig1} and \ref{fig3}). However, for $\Delta /\hbar
\omega =1.1$ in Fig.\ \ref{fig6}(b), the $S$ and $A$ states have an apparent
exchange of behavior in the $F$ field, which is also exhibited for all $\Delta
/\hbar \omega >1$ (as in Fig.\ \ref{fig4}). This apparent switching is due to
the mixing (and level repulsion) seen near $F\simeq 0$, and the fact that the
labels are always obtained for the $F=0$ limit, where the AC field is
non-existent, and the spatial symmetry is well defined.\footnote{Notice that
for $\Delta/\hbar\omega=odd$, the quasienergies show no gap at $F=0$, but split
quadratically away, similar to an anticrossing behavior. This however, does not
yield dynamic localization and agree with a set of solutions to Eq.\
(\ref{t14}).} Notice, interestingly, that the related AC Stark effect in atoms
and superlattice systems\cite{19a} occurs for $\Delta
>\hbar \omega $, but this effect is somewhat different.

\subsection{Incomplete dynamic localization}

Comparing the cases analyzed so far, one promptly notices that the `degree'
of dynamic location changes with the ratio $\Delta /\hbar \omega $. In the
limit $\Delta /\hbar \omega $ $\ll 1$, we saw that the $l_{i}$ values tend
to the zeros of the Bessel function, and that the probability of staying
indefinitely in the left well is approximately 1 at all times (full
localization). For $\Delta /\hbar \omega =0.5$, Fig.\ \ref{fig1}, the
dynamic localization points occur for somewhat smaller values of the AC
field, and the probability of staying fully in the left well is smaller
(poorer localization). In fact, the particle is able to increasingly leak
out to the right well for lower frequencies, even for strong AC field
amplitudes.

\begin{figure}[tbp]
\includegraphics*[width=8.6cm]{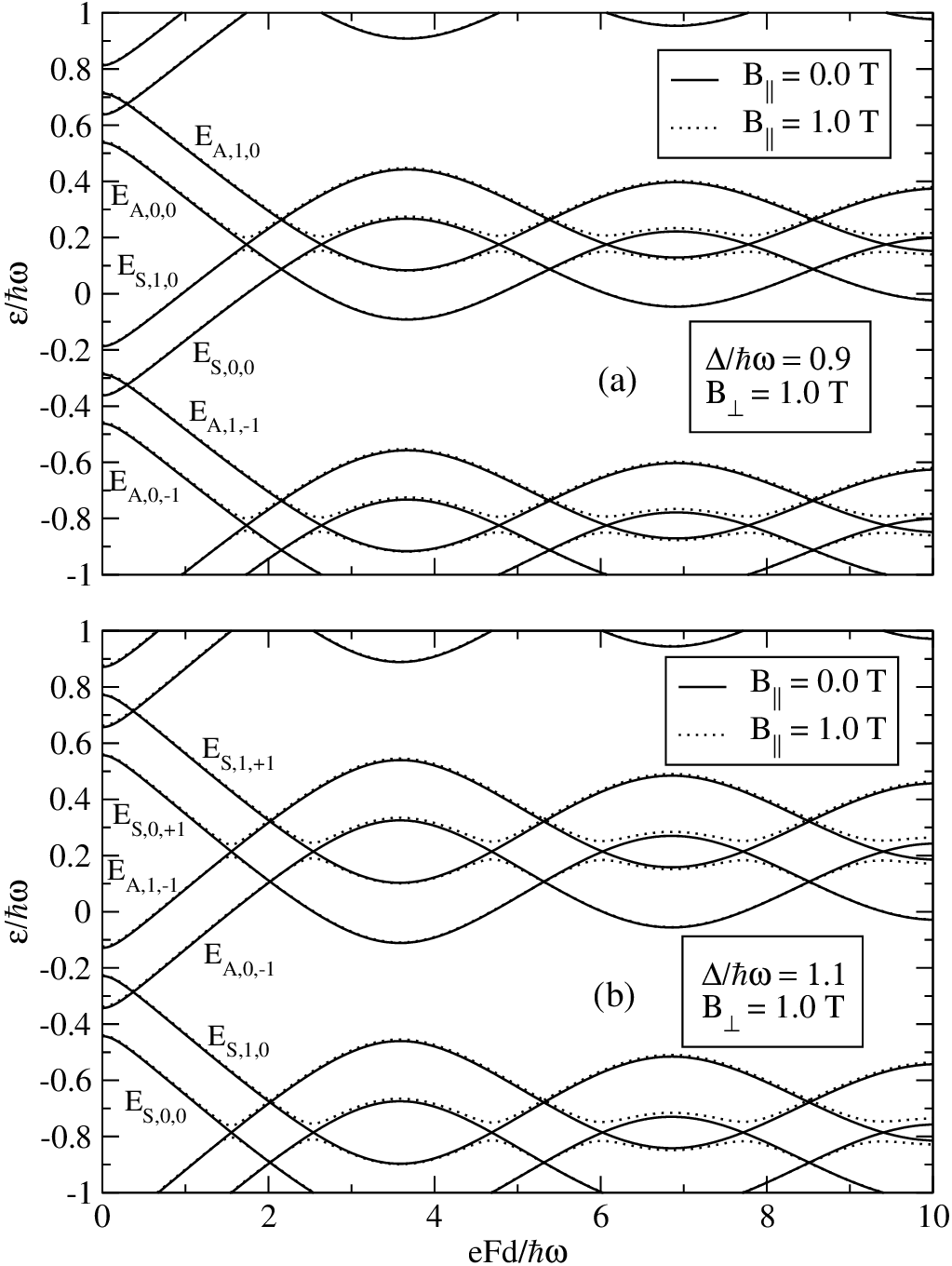}
\caption{Two Brillouin zones of quasienergies (in units of $\hbar \protect%
\omega $) as function of $eFd/\hbar \protect\omega $ (intensity of AC
field), for: (a) $\Delta /\hbar \protect\omega =0.9$ and (b) $\Delta /\hbar
\protect\omega =1.1$, for $B_{\perp }=1.0$~T. Notice, despite the similarity
of the pictures, that the labels $E_{s,n,m}$ are completely different. }
\label{fig6}
\end{figure}

\begin{figure}[tbp]
\includegraphics*[width=8.6cm]{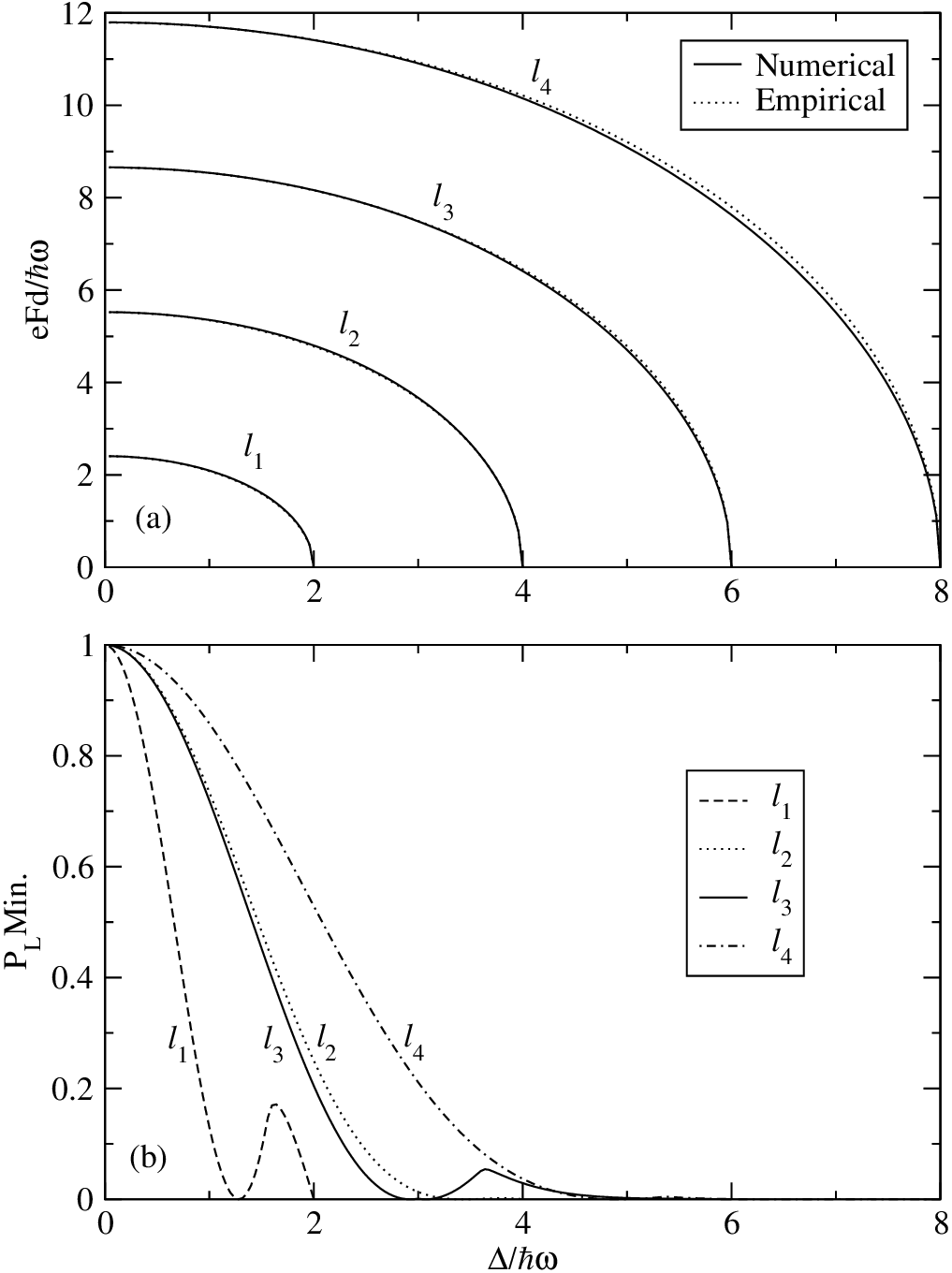}
\caption{(a) Exact crossing points $l_{i}$ in the first Brillouin zone of
quasienergies which yield localization, as function of the ratio $\Delta
/\hbar \protect\omega $. Full lines are from our numerical calculation;
dotted lines are empirical fits given by Eq.~(\ref{empiric}). (b)
Corresponding minima in the probability to remain in the left well for these
crossing points. Larger $P_{L}$ minima indicate better dynamical
localization, such as for $\Delta /\hbar \protect\omega \lesssim 1$.}
\label{fig7}
\end{figure}

A more complete analysis of this behavior is shown in Fig.\ \ref{fig7}. The
full line in \ref{fig7}(a) shows how the dynamic localization points change
with the ratio $\Delta /\hbar \omega $. We have numerically computed this
function by looking for the crossing points in the first quasienergy
Brillouin zone that produce localization behavior ($l_{i}$). An interesting
result, as described above, is that the tilted magnetic field has no
discernible effect on these points, so that this result is identically the
same as for a simple two level problem. Notice that increasing the $\Delta
/\hbar \omega $ ratio not only decreases the $F$ values needed for
localization, but eventually suppress a given $l_{j}$, at even values $%
\Delta /\hbar \omega =2j$, where the unperturbed photon replicas would have
degeneracy points (see discussion after Eq.\ (\ref{t14})). The odd ratios do
not yield level crossing and localization, as the $S_{p}$ parity symmetry of
nearby states is \emph{the same} ($E_{S,0,0}$ and $E_{A,0,-1}$, for example,
for $\Delta \simeq \hbar \omega $), and states in fact mix (anticross) as we
discuss above in reference to Fig.\ \ref{fig6}. We should emphasize that
the high frequency results discussed in the literature and yielding dynamic
localization at the zeroes of the Bessel functions, agree identically with
our numerical results. Similarly, the weak field limit yields $l_{i}$ points
at integer values of $\Delta /\hbar \omega $, as given by the degeneracy
condition, Eq.\ (\ref{t14}). That equation, however, does yield an
appropriate estimate of the $l_{i}$ only in the extreme limits of field and
frequency.

Inspection of Fig.\ \ref{fig7}(a) suggests parabolic curves. In fact, in a
fully empirical way we use
\begin{equation}
l_{j}=\sqrt{x_{j}{}^{2}-\left( \frac{x_{j}}{2j}\frac{\Delta }{\hbar \omega }%
\right) ^{2}},  \label{empiric}
\end{equation}%
where $x_{j}$ is the $j^{th}$-zero of the Bessel function, $J_{0}(x_{j})=0$.
This simple fit to the end points describes surprisingly well the curves
over the whole range, as one can see in the dotted lines in Fig.\ \ref{fig7}%
(a). The fit is so good, that it is basically indistinguishable for the
first localization point $l_{1}$, and it only departs noticeably for $l_{3}$
and $l_{4}$.

In addition to the shift in position of the localization points $l_{j}$, and
as noticed in the discussion before, we find that the value of $\Delta
/\hbar \omega $ also changes the \emph{degree} of localization of the
particle. To further study this point, we have arrived at a quantitative
measure of how good the localization is, by looking at the minimal
probability of remaining in the left well during a periodic oscillation of $%
P_{L}(t)$. For long times, a large minimal value of $P_{L}$ illustrates that
the localization is `good'. A small value of this quantity indicates that
the particle is able to jump or tunnel to the right well with a large
probability, even if the particle eventually returns fully to the original
the localization is then `poor'. Figure\ \ref{fig7}(b) presents the resulting
presents the resulting minimal $P_{L}$ for different dynamical localization
points $l_{j}$. These curves have some surprising behavior: (a) Increasing
the ratio $\Delta /\hbar \omega $ produces a rapid deterioration of the
dynamic localization points for all $l_{j}$. In fact, for $\Delta /\hbar
\omega >5$, all localization points shown exhibit such poor localization
(minimal $P_{L}\simeq 0$), that the time-dependent oscillations are
basically non-localized (even though the $F$ field values are still somewhat
large, Fig.\ \ref{fig7}(a)). (b) For a range of $\Delta /\hbar \omega $, the
second crossing point $l_{2}$ is in fact better localized than $l_{3}$,
unlike the situation at high frequency. (c) For $l_{1}$ and $l_{3}$ (but not
for $l_{2}$ and $l_{4}$), there is a sudden `\emph{revival}' of
localization, as the `bumps' in the those curves indicate. This `revival'
behavior is not intuitive and its source not yet understood. We have
attempted to link this to the details of the quasienergy spectra (such as
possible degeneracies or level anticrossings). We have not been able to
identify a clear correlation.

\begin{figure}[tbp]
\includegraphics*[width=8.6cm]{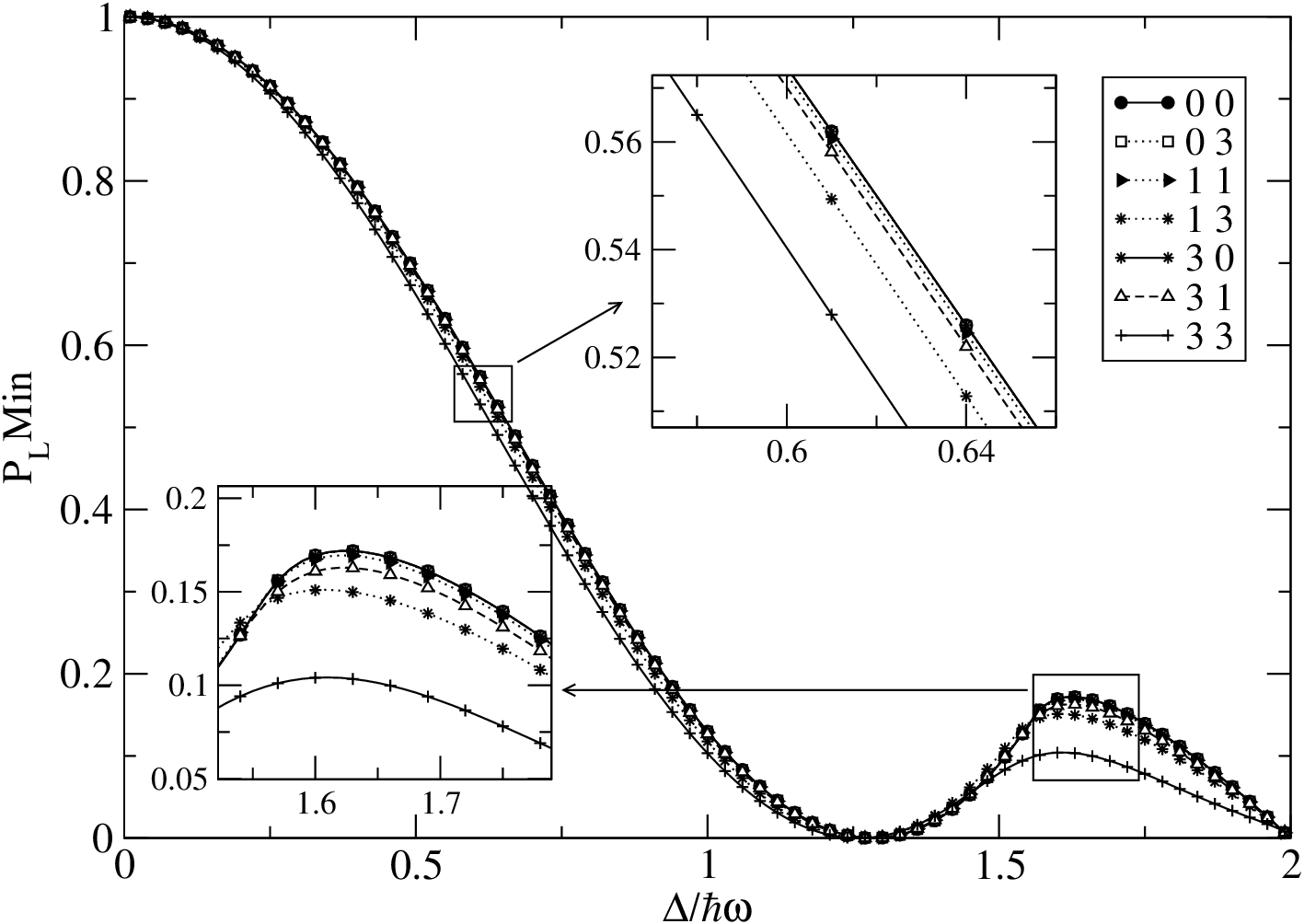}
\caption{Minimal probability to remain in the left well for the first
crossing points $l_{1}$, for several values of magnetic field. The legend
indicates the two magnetic field components; the first number (left) refers
to the intensity of $B_{\perp }$ and the second (right) to $B_{\parallel }$
in Tesla. Insets show magnification of different curve regions, for clarity.}
\label{fig8}
\end{figure}

We have also analyzed to what extent the degree of localization is affected
by other parameters, such as the in-plane field, $B_{\parallel }$. Figure %
\ref{fig8} illustrates the minimal $P_{L}$ for different values of the
magnetic field. We find that to a great extent, the field (both in-plane and
perpendicular) produces no change in the main drop of the $l_{1}$ curve.
However, the height and peak position of the `revival feature' are more
noticeably changed by the magnetic field. Although we have computed these
curves for a wide variety of total magnetic field amplitudes and/or angles,
we find that the curve is not drastically modified, and at most yields a
suppression of this revival feature by a factor of nearly two. Notice,
however, that the changes appear only when both components of the field are
present, and are larger for the larger total field attempted. Again, the
quasienergies show no evident feature that one can associate with these
changes, and its precise elucidation remains a challenge for future work.

\section{\label{Conclusion}Conclusions}

We have analyzed a double quantum well system in the presence of a tilted
magnetic field, and a strong oscillating electric field. We find, as other
authors, that for high frequency, the otherwise tunneling particle is
localized in one of the two wells at certain values of the electric field
amplitude. We have however, studied this system over the entire regime of
frequencies and field amplitudes. We find, in agreement with recent
analytical work, that the electric field at which localization occurs
changes monotonically down with lower frequency and that this trend
continues far away from the perturbative regime of $\Delta /\hbar \omega \ll
1$. We find however, that the degree of localization does not change in a
monotonic fashion, and even shows a certain degree of recovery or revival,
for lower frequencies. This rather involved behavior with the different
system parameters is likely to produce a number of experimentally observable
consequences. For example, the varying degree of charge localization for
different frequency values in a given structure would yield variations in
the generation power of THz radiation. The tunability with $B_{\Vert }$, on
the other hand, would produce entirely different frequencies in the
generation of radiation, away from the anticipated driving harmonics. This
behavior would then contribute significantly to the already rich assortment
of different phenomena observed in these systems.

\begin{acknowledgments}
J.\ M.\ Villas-B\^{o}as and P.\ H.\ Rivera acknowledge the financial support of
Funda\c{c}\~{a}o de Amparo \`{a} Pesquisa do Estado de S\~{a}o Paulo
(FAPESP). We acknowledge the partial support from US DOE grant no.
DE-FG02-91ER45334.
\end{acknowledgments}

\bibliography{Paper}

\end{document}